\documentclass[showpacs,a4paper]{revtex4}
\usepackage{graphicx}

\begin{document}

\title{Microwave-induced coupling of superconducting qubits}

\author{G.~S.~Paraoanu}\email{paraoanu@phys.jyu.fi}

\affiliation{NanoScience Center and Department of Physics, University of Jyv\"askyl\"a,
P.O.~Box 35 (YFL), FIN-40014 University of Jyv\"askyl\"a, Finland}

\affiliation{Institute for Quantum Optics and Quantum Information, ICT-Geb\"aude, Technikerstra\ss e 21a, A-6020 
Innsbruck, Austria}

\begin{abstract}
We investigate the quantum dynamics of a system of two coupled superconducting qubits under microwave 
irradiation. We find that, with the qubits operated at the charge co-degeneracy point,
the quantum evolution of the system can be described by a new effective Hamiltonian which has the form of two 
coupled qubits with tunable coupling between them. This Hamiltonian can be used for experimental tests on 
macroscopic entanglement and for implementing quantum gates.
\end{abstract}


\maketitle


A significant interest in the quantum coherence properties of various superconducting devices
has been manifest in the last years following the successful demonstration of superpositions
of charge and flux based macroscopic quantum states \cite{coherence}. It was immediately realized that, when 
supplemented with
appropriate read-out components and protocols, these devices qualify
as candidates for quantum bits in future quantum computing architectures \cite{reviews}. Indeed, several types 
of superconducting
qubits such as phase qubits \cite{phase}, charge qubits \cite{charge}, charge-phase qubits \cite{vion}, and flux 
qubits \cite{irinel} have been operated since.

For systems of two superconducting qubits, a desirable feature to have in order to implement two-qubit gates is 
tunable coupling, and several schemes using for instance variable electrostatic transformers \cite{averin}, the 
dynamic inductance of a dc SQUID \cite{clark},
or resonator circuits \cite{resonatorcoupling} have been proposed. These ideas have not yet been fully tested 
experimentally: instead, fixed coupled qubits have been studied as a preliminary step, and a few significant 
results have already been reported, such as signatures of entanglement in coupled phase qubits 
\cite{entanglement}, and a protocol for implementing a C-NOT gate \cite{cnot}. However, it has been recognized 
that these protocols will not be easily scalable because they manipulate the qubits far from the degeneracy 
points, where decoherence is strong (for single charge qubits, when operated off-degeneracy, the dephasing times 
are of the order of only a few  hundred picoseconds \cite{echoes}, due mostly to 1/f noise).
The solution to this problem in the case of single qubits \cite{vion} is to keep the qubit at the degeneracy 
point during the quantum gate, and to move away from this point only during the measurements; this strategy has 
been successfully implemented for flux qubits as well \cite{irinel}.
But for two qubits the requirements are rather contradictory: on one hand we would like to have the qubits 
operated at the special, low-decoherence degeneracy points; on the other hand, adding up the coupling will 
remove them from these points. More recent proposals have attempted to solve this problem by employing NMR-style 
strategies \cite{rigetti}, or by using a superconducting circuit that allows modulation of the coupling 
\cite{bertet} or act as the vibration mode of two trapped ions \cite{nori}.

In this paper we show that it is possible to create entanglement and quantum gates in
a system of two qubits with fixed coupling, irradiated with a monochromatic off-resonance microwave field and 
biased at the co-degeneracy point. As a result, the proposed quantum circuit satisfies both of the requirements 
above: it is insensitive to noise due to fluctuations of the external parameters ({\it e.g.} 1/f noise), and it 
can be mapped into a system of two qubits with tunable coupling. The scheme is therefore minimal
from the point of view of the number of on-chip circuit elements required and can be realized
immediately without any major change in the existing qubit experimental setups.
Also, in contradistinction with the fast pulse method of \cite{charge,cnot} this
technique does not rely on high-precision microwave electronic equipment, therefore, from an experimental point 
of view, could
be regarded as more reliable and relatively inexpensive.

\begin{figure}[htb]
\includegraphics[width=75truemm]{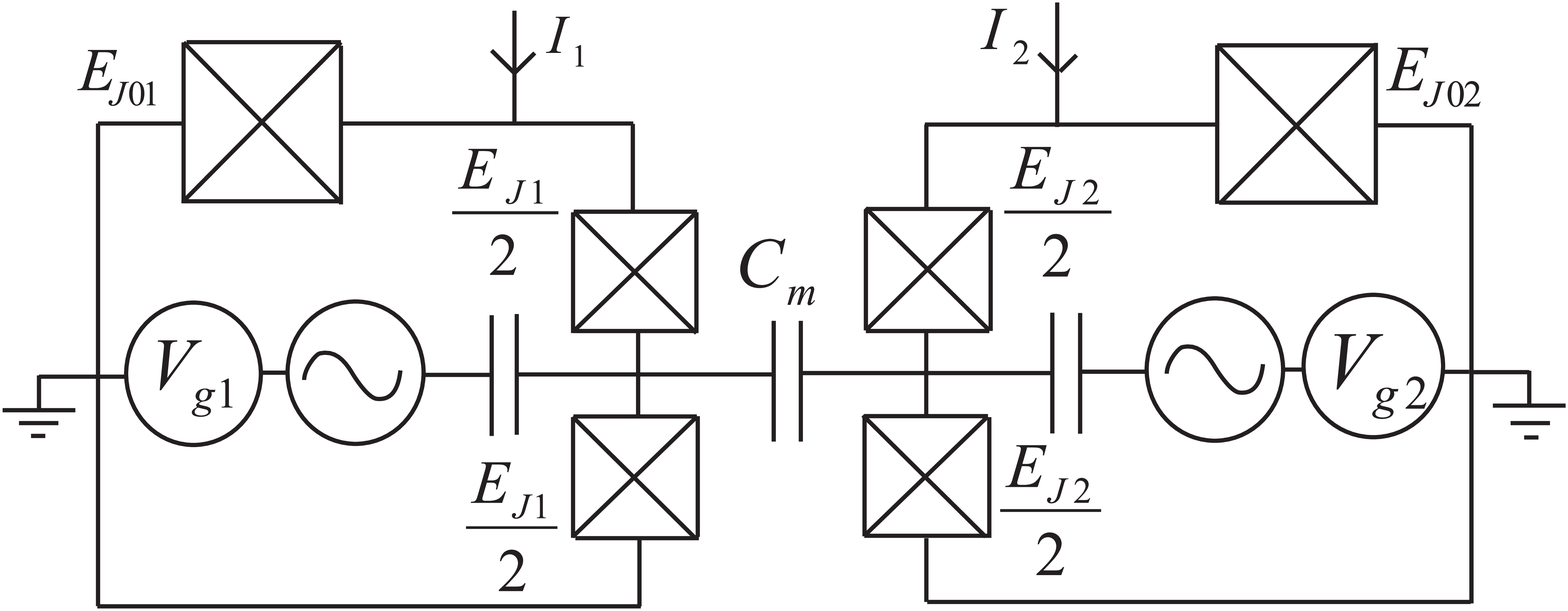}
\caption{Schematic of the circuit: two capacitively coupled Quantronium circuits.}
\label{schematic}
\end{figure}

We do all the calculations for the case of coupled Quantronium circuits \cite{vion}, with the observation
that the results can be easily adapted to almost any other qubit species.
Let us  start with the Hamiltonian for two capacitively coupled charge qubits \cite{cnot},
\begin{eqnarray}
H&=&E_{C1} (n_{g1}-n_{1})^{2} - E_{J1}\cos{\varphi_1}
+ E_{C2} (n_{g2}- n_{2})^{2} \nonumber \\
& &- E_{J2}\cos{\varphi_2} + E_{m}(n_{g1} - n_{1})(n_{g2}-n_{2}), \label{initial}
\end{eqnarray}
with $E_{C1,2}\approx 2e^{2}/C_{\Sigma 1,2}$, $E_{J1,2}$ the standard charging and Josephson energies 
(Fig.\ref{schematic}) for each split Cooper pair box
($C_{\Sigma 1,2}$ are predominantly given, for each qubit, by the sum of the corresponding island-to-lead 
capacitances) and $E_{m}\approx 4e^{2}C_{m}/C_{\Sigma 1}C_{\Sigma 2}$ ($C_{m}\ll C_{\Sigma 1,2}$ is the coupling 
capacitance).

The condition of insensitivity to fluctuations in the external parameters
is satisfied automatically if the operating point of the qubits is fixed
at $n_{g1}=n_{g2}=1/2$ -  the so-called co-degeneracy point - where the
eigenvalues of the Hamiltonian are first-order independent
of fluctuations in $n_{g1}$ and $n_{g2}$.
At the codegeneracy point, the 2-qubit Hamiltonian Eq.(\ref{initial}) written in the
eigenbasis $|\uparrow > = \left( |0> +  |1> \right)/\sqrt{2}$,
$|\downarrow > = \left( |0> -  |1> \right)/\sqrt{2}$
of each qubit -- considering the approximation of large charging energies
in which we can restrict ourselves to the subspace spanned by
$\{|n_{1}>\} = \{|0>,|1>\}$, $\{|n_{2}>\}=\{|0>,|1>\}$ --
has the form
\[ H = \frac{1}{4}(E_{C1} + E_{C2}) + \frac{1}{4}E_{m}\sigma^{x}_{1}\sigma^{x}_{2}
-\frac{1}{2}E_{J1}\sigma^{z}_{1}-\frac{1}{2}E_{J1}\sigma^{z}_{2}. \]
This Hamiltonian
can be diagonalized
by introducing the matrix
\begin{equation}
M_{\pm} = \left( \begin{array}{cc} \cos\theta_\pm /2 & -\sin\theta_\pm /2 \\ \sin \theta_\pm /2 &
\cos \theta_\pm /2 \end{array}\right),
\end{equation}
where $M_{+}$ acts on the subspace spanned by $\{|\uparrow\uparrow >, |\downarrow\downarrow >\}$
and $M_{-}$ acts on the subspace $\{|\downarrow\uparrow >, |\uparrow\downarrow >\}$.
The angle $\theta_\pm$ is defined by
\begin{equation}
\tan\theta_\pm = -\frac{E_m}{2(E_{J2} \pm E_{J1})}.
\end{equation}
The eigenbasis for the coupled qubits is
\begin{eqnarray}
\left(\begin{array}{c}|uu> \\ |dd> \end{array}\right)&=& M_{+}^{-1}\left(\begin{array}{c}|\uparrow\uparrow >
\\ |\downarrow\downarrow > \end{array}\right), \label{unu1} \\
\left(\begin{array}{c} |du> \\ |ud> \end{array}\right)&=&M_{-}^{-1}\left(\begin{array}{c} |\downarrow\uparrow >
\\ |\uparrow\downarrow > \end{array}\right).\label{doi2}
\end{eqnarray}

The energy levels of the 4-level system are defined by the quantities $\xi$ and $\epsilon$,
\begin{eqnarray}
\xi &=& \frac{1}{2}\sqrt{(E_{J1} + E_{J2})^{2} + (E_{m}/2)^2}, \\
\epsilon &=& \frac{1}{2}\sqrt{(E_{J1} - E_{J2})^{2} + (E_{m}/2)^2}.
\end{eqnarray}
It is also useful to introduce the notation $\hbar \Omega = \xi + \epsilon$ and $\hbar \nu = \xi - \epsilon$.
Since we will need to use Pauli operators also with respect to the new fixed-coupling eigenbasis
Eq. (\ref{unu1},\ref{doi2}), we adopt the convention that $\sigma^{\{x,y,z\}}_{\{1,2\}}$ refer to the original 
qubits, and $\sigma_{x,y,z}$ together with the
tensorial product $\otimes$ correspond to the basis Eq. (\ref{unu1},\ref{doi2}).
For example, in the fixed-coupling eigenbasis the Hamiltonian describes two uncoupled qubits,
$H = - \hbar (\nu \sigma_{z}\otimes I + \Omega I\otimes \sigma_{z})/2$.

Consider now the case of an excitation produced by irradiating the qubits with a monochromatic microwave 
radiation of angular frequency $\omega$,
$n_{g1} = 1/2 + w_{1}\cos\omega t$ and $n_{g2} = 1/2 + w_{2}\cos\omega t$.
The quantities $w_{1}$ and $w_{2}$ are the amplitudes of the radiation: experimentally, they can be adjusted 
relatively fast and
independently for each qubit, by mixing the continuous microwave with a tunable signal
from a pulse pattern generator or an arbitrary waveform generator.
The Hamiltonian is then
\begin{eqnarray}
H &=& \frac{1}{4}(E_{C1} + E_{C2}) +
E_{C1}w_{1}\cos\omega t\sigma^{x}_{1}
+ E_{C2}w_{2}\cos\omega t\sigma^{x}_{2} + \nonumber \\
& &
 \frac{1}{4}E_{m}\sigma^{x}_{1}\sigma^{x}_{2}
-\frac{1}{2}E_{J1}\sigma^{z}_{1}-\frac{1}{2}E_{J1}\sigma^{z}_{2}.\label{asta}
\end{eqnarray}

We expand at any time $t$ the state vector by separating the eigenenergies $\pm \xi, \pm \epsilon$ of the
four states $|uu>,|dd>$ and $|du>,|ud>$,
\begin{equation}
\Psi (t)= e^{\frac{i}{\hbar}\xi t} c_{uu}(t) |uu >  + e^{\frac{i}{\hbar}\epsilon t}c_{du}(t) |du > + 
e^{-\frac{i}{\hbar}\epsilon t}c_{ud}(t) |ud > + e^{-\frac{i}{\hbar}\xi t}c_{dd}(t) |dd >.
\end{equation}
Inserting this expansion into the Schr\"odinger equation with Hamiltonain Eq. (\ref{asta}), one can notice that 
it is
possible to perform a rotating-wave approximation by neglecting fast-oscillating terms
(containing the frequencies $\omega + \Omega$ and $\omega + \nu$; below we will neglect the Bloch-Siegert shifts 
\cite{bs} which become important only for higher values of the microwave amplitude).

If we define the detunings of the external microwave radiation from the two-qubit frequencies $\delta = \omega - 
\nu$ and $\Delta = \omega - \Omega$, we obtain
the system of equations
\begin{eqnarray}
i\hbar \frac{d}{dt} \tilde{c}_{uu} &=& \frac{\hbar}{2}(\delta + \Delta ) \tilde{c}_{uu} + 
2T^{(1)}_{+}\tilde{c}_{du} + 2T^{(2)}_{+}\tilde{c}_{ud} ,\\
i\hbar \frac{d}{dt} \tilde{c}_{du} &=&  2T^{(1)}_{+}\tilde{c}_{uu} + \frac{\hbar}{2}(-\delta + \Delta ) 
\tilde{c}_{du}  + 2T^{(2)}_{-}\tilde{c}_{dd} ,\\
i\hbar \frac{d}{dt} \tilde{c}_{ud} &=&  2T^{(2)}_{+}\tilde{c}_{uu} + \frac{\hbar}{2}(\delta -\Delta 
)\tilde{c}_{ud} + 2T^{(1)}_{-}\tilde{c}_{dd} ,\\
i\hbar \frac{d}{dt}\tilde{c}_{dd} &=&  2T^{(2)}_{-}\tilde{c}_{du} + 2T^{(1)}_{-}\tilde{c}_{ud} + 
\frac{\hbar}{2}(-\delta -\Delta ) \tilde{c}_{dd},
\end{eqnarray}
where we introduced the notations
\begin{eqnarray}
&&T^{(1)}_{\pm} = \frac{E_{C1}}{2}w_{1}\cos\frac{\theta_{+} -\theta_{-}}{2} \pm 
\frac{E_{C2}}{2}w_{2}\sin\frac{\theta_{+} + \theta_{-}}{2}, \nonumber \\
&&T^{(2)}_{\pm} = \frac{E_{C2}}{2}w_{2}\cos\frac{\theta_{+} +\theta_{-}}{2} \pm 
\frac{E_{C1}}{2}w_{1}\sin\frac{\theta_{+} - \theta_{-}}{2}, \nonumber
\end{eqnarray}
and the following substitutions have been used:
$c_{uu}(t)=\exp[i(\delta + \Delta ) t/2]\tilde{c}_{uu}(t)$,
$c_{du}(t)=\exp[i(-\delta + \Delta ) t/2]\tilde{c}_{du}(t)$,
$c_{ud}(t)=\exp[i(\delta - \Delta ) t/2]\tilde{c}_{ud}(t)$,
$c_{dd}(t)=\exp[i(-\delta - \Delta ) t/2]\tilde{c}_{dd}(t)$.
This is a quantum evolution
governed by a time-independent effective Hamiltonian
\begin{eqnarray}
H_{eff} &=& \left[\frac{\hbar\delta}{2}\sigma_{z}+ \frac{E_{C1}w_{1}}{2}\cos\frac{\theta_{+}-\theta_
{-}}{2}\sigma_{x}\right]\otimes I
+ I\otimes \left[\frac{\hbar\Delta}{2}\sigma_{z}+ \frac{E_{C2}w_{2}}{2}\cos\frac{\theta_{+}+\theta_
{-}}{2}\sigma_{x}\right] \nonumber \\
& & + \frac{E_{C1}w_{1}}{2}\sin\frac{\theta_{+} -\theta_{-}}{2}\sigma_{z}\otimes\sigma_{x} + 
\frac{E_{C2}w_{2}}{2}\sin\frac{\theta_{+} + \theta_{-}}{2}\sigma_{x}\otimes\sigma_{z}\label{thisone}
\end{eqnarray}

The Hamiltonian  Eq. (\ref{thisone}) is the central result of this paper; it describes a set of two coupled 
qubits of states $|u>, |d>$ with the coupling controlled
by the radiation intensities $w_1$ and $w_2$.
The elements of this Hamiltonian, namely $\sigma_{z}\otimes I$, $I\otimes \sigma_{z}$, $\sigma_{x}\otimes I$, 
$I\otimes \sigma_{x}$, $\sigma_{z}^{1}\otimes\sigma_{x}^{2}$, and $\sigma_{z}^{2}\otimes\sigma_{x}^{1}$ span the 
whole su(4) Lie algebra \cite{geo}.
The entangling properties of this type of Hamiltonian
have not been studied before in the quantum computing community and no ready-made analytical recipe exists for 
the problem of
generating a given two-qubit gate. We approach this problem numerically.

A typical experimental situation will be $E_{J1}\gg E_{m}$, $E_{J2}\gg E_{m}$, therefore $|\theta_{+}|\approx 
0$. Also, we want to avoid an entangling dynamics between the qubits in the absence of quantum gate driving, 
therefore we impose the quasi-separability condition
$E_{m}\ll |E_{J1}-E_{J2}|$ (a unitary transformation in the 4-state basis $|uu>,|du>,|ud>,|dd>$
will be approximately the same when written in the original qubit basis $|\uparrow\uparrow>, 
|\downarrow\uparrow>, |\uparrow\downarrow>, |\downarrow\downarrow>$). With these approximations, the Hamiltonian 
Eq.(\ref{thisone}) reads
\begin{eqnarray}
H_{eff} &=& \left[\frac{\hbar\delta}{2}\sigma_{z}+ \frac{E_{C1}w_{1}}{2}\sigma_{x}\right]\otimes I + I\otimes 
\left[\frac{\hbar\Delta}{2}\sigma_{z}+ \frac{E_{C2}w_{2}}{2}\sigma_{x}\right]\nonumber \\
& &+ \frac{E_{m}}{8 (E_{J2}-E_{J1})}\left(E_{C1}w_{1}\sigma_{z}\otimes \sigma_{x} - E_{C2}w_{2}\sigma_{x}\otimes 
\sigma_{z}\right).
\end{eqnarray}
We now see that in the presence of the driving microwave field, the relatively small value of $E_{m}$ is 
compensated by the field intensity, and controlled entanglement
becomes possible as the Larmor frequency of one
qubit is modulated by the transversal part of the Rabi oscillations of the other one.

We have calculated numerically the evolution of the amplitudes $\tilde{c}$
for different values of the Quantronium parameters.  Two interesting particular cases emerge.
Let us for simplicity neglect $\theta_{+}\approx 0$ then take
$\delta = -\Delta >0$ ($\hbar\omega = \hbar (\nu + \Omega)/2 =\xi$) and $E_{C1}w_{1}=E_{C2}w_{2}=\hbar W$. The
effective Hamiltonian has then one eigenvalue zero and the other three given by the solutions of the third order 
equation $\lambda^{3}-\lambda(\delta^{2}+W^{2})+\delta W^{2}\sin\theta_{-} =0$.
Consider first the situation $\delta \gg W$.
The solutions for the finite eigenvalues are found
approximately $\lambda_{1,2} \approx \pm \delta - (W^{2}/2)\delta^{-1}\sin\theta_{-}$ and $\lambda_{3}\approx 
W^{2}\delta^{-1}\sin\theta_{-}$. The evolution can be approximated by $\tilde{c}_{uu} = [1 + \exp (-i\lambda_{3} 
t)] /2$, $\tilde{c}_{dd} = [-1 + \exp (-i\lambda_{3} t)] /2$, $\tilde{c}_{du}=\tilde{c}_{ud}=0$, and the 
concurence
assumes a remarkable simple form, ${\cal C} = 2 |c_{uu}c_{dd} - c_{du}c_{ud}| = |\sin \lambda_{3} t|$. In this 
case we note that the buildup of a probability amplitude on the state
$|dd>$ is due to a coherent effect similar to that which produces dark states for $\Lambda$ - atoms. There is
no matrix element between the initial state $|uu>$ and $|dd>$; instead, the atoms are transferred coherently
to $|uu>$ through $|ud>$ and $|du>$, which have population zero due to destructive interference of the amplitude 
probabilities. This results at times $\lambda_{3}^{-1}( \pi /2 + n\pi )$ in the creation of a maximally 
entangled state of the form
\[
\frac{1}{2}[1 + i(-1)^{n+1}]|uu> + \frac{1}{2}[-1 + i(-1)^{n+1}]|dd> .
\]
The time required is of the order of $\lambda_{3}^{-1}$, and it gets larger for smaller and smaller microwave 
power.

\begin{figure}[htb]
\includegraphics[width=75truemm]{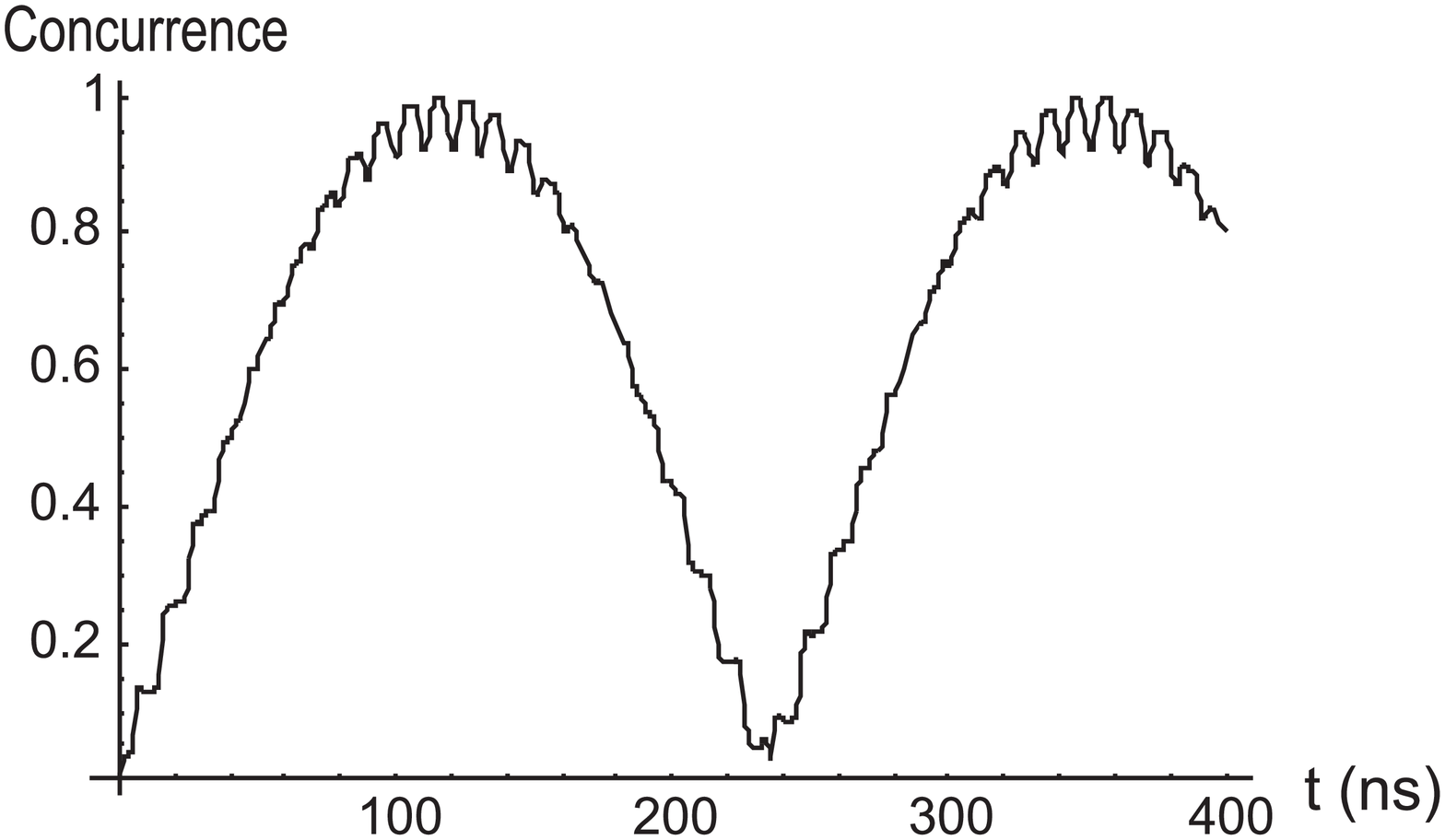}
\caption{Concurrence for the case $W\gg \delta$. The parameters for this figure are $W=2\pi \times 96.3$MHz,
$\delta = -\Delta = 2\pi \times 15.2$ MHz, $\theta_{+}\approx 0$, and $\theta_{-}=-0.29$}
\label{concurrenceWlarge}
\end{figure}

The second case is $W\gg \delta$. In this case, the eigenvalues can be still determined
approximately, $\lambda_{1,2}= \pm W -(\delta /2)\sin\theta_{-}$ and $\lambda_{3}=\delta\sin\theta_{-}$, but
the evolution of the coefficients $c_{uu,du,ud,dd}$ is not so simple anymore. Still, we have verified
numerically that also in this limit the concurrence assumes a rather simple form (Fig.\ref{concurrenceWlarge}), 
but this
time the oscillation period is not related only to $\lambda_{3}$. The state at ${\cal C}=1$
has components on all four vectors $|uu>,|du>,|ud>,|dd>$, which tend to oscillate on a timescale of the order of 
$W^{-1}$;
remarkably, this evolution conspires to give a concurrence of a simple form, as shown in the figure.

In principle, one can measure
these amplitudes (details at the end of the paper), and compare the results with the
theory.
These experiments are important because they are simple to realize (microwave amplitude is kept constant)
and the theoretical prediction is an oscillation of the concurrence between the extreme values 0 and 1, 
therefore
the comparison with experimental data (as well as the extraction of 2-qubit dephasing times) is straightforward.

\begin{figure}[htb]
\includegraphics[width=75truemm]{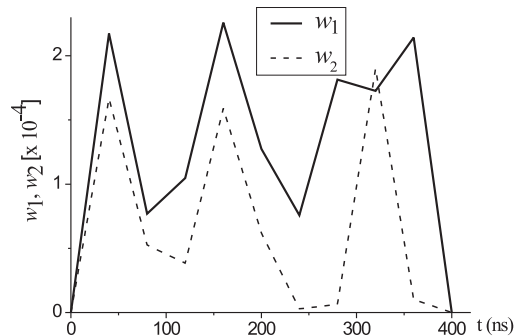}
\caption{Paramers $w_{1}$ and  $w_{2}$ for CNOT ($E_{C1}=605$GHz, $E_{C2}=454$GHz, $\delta = -\Delta=-19.2$ MHz, 
$\theta_{+}\approx 0$, $\theta_{-}=-0.29$).}
\label{cnotannealing}
\end{figure}

We now address the problem of implementing quantum gates numerically. We first notice (and we also checked 
numerically) that local unitary transformations (single-qubit gates) can be realized simply by tuning the 
incoming microwave
in resonance with either one of the two qubits and using relatively low power.
For two-qubit gates we have to find appropriate excursions in the parameter space
$\{w_{1}, w_{2}\}$ such that the result of the evolution has the same Makhlin invariants $G_{1}$
and $G_{2}$ as
CNOT \cite{numerics}; we therefore have to perform a numerical search for the minimum of
the expression $|G_{1}(t)|^{2} + |G_{2}(t)-1|^{2}$.
Our search method is based on simulated annealing \cite{harri};
for example in the case of equal detunings $\delta = -\Delta$ one possible control parameter
sequence $\{w_{1},w_{2}\}= \{w_{1}(t),w_{2}(t)\}$ is presented in
Fig. \ref{cnotannealing}.

{\it Measurements.} We consider a measurement scheme (Fig.\ref{schematic}) in which each of the two qubits is 
shunted by
large current-biased junctions. As in the single-qubit Quantronium, the bias currents
$I_1$ and $I_2$ will be kept to zero during the two-qubit gate. Next, currents are raised adiabatically to a 
value close to the
critical current of the large junctions, and a switching event in a chosen time interval
is recorded
or not. Upon performing this experiment a large number of times, switching probabilities
- as functions of the quantum state of the qubit - can be determined experimentally.
In the approximation $E_{m}\ll |E_{J1}-E_{J2}|$ ($|uu > \approx |\uparrow \uparrow >$, $|du > \approx 
|\downarrow\uparrow >$, etc.), and the equations of motion for the
 macroscopic phase differences across the large read-out junctions separate. Therefore
 one can introduce independent switching rates $\Gamma_{1}^{\uparrow}$ and $\Gamma_{2}^{\downarrow}$ defined, as 
in the
 single-qubit case \cite{vion,paraoanu}, at two bias currents $I_{1}$ and $I_{2}$, where the sensitivity of the 
measurement is maximal.
Let us now imagine a switching current experiment in which the large junctions are biased at $I_{1}$ and $I_{2}$
for times $\tau_{1}$ and $\tau_{2}$ respectively. The experimentalist can measure $P_{{\rm yes,yes}}$, $P_{{\rm 
yes,no}}$, $P_{\rm{no,yes}}$,
and $P_{\rm{no,no}}$, the probabilities that both junctions have switched, that the first has switched while the 
second did not, etc.
Using for instance the formalism described in \cite{paraoanu} one can show that these probabilities are given by
\begin{eqnarray}
P_{\rm{no,no}}&=&|c_{\uparrow\uparrow}|^{2}e^{-\Gamma^{\uparrow}_{1}\tau_{1}}e^{-\Gamma^{\uparrow
}_{2}\tau_{2}} + |c_{\downarrow\uparrow}|^{2}e^{-\Gamma^{\downarrow}_{1}\tau_{1}}e^{-\Gamma^{\uparrow}_{2}
\tau_{2}} \nonumber \\ & &
+|c_{\uparrow\downarrow}|^{2}e^{-\Gamma^{\uparrow}_{1}\tau_{1}}e^{-\Gamma^{\downarrow}_{2}\tau_{2}}
+|c_{\downarrow\downarrow}|^{2}e^{-\Gamma^{\downarrow}_{1}\tau_{1}}e^{-\Gamma^{\downarrow}_{2}\tau_{2}}; 
\nonumber \\
P_{\rm{yes,no}}&=&-P_{\rm{no,no}} + (|c_{\uparrow\uparrow}|^{2} + 
|c_{\downarrow\uparrow}|^{2})e^{-\Gamma^{\uparrow}_{2}\tau_{2}} + \nonumber \\ & &
(|c_{\uparrow\downarrow}|^{2} + |c_{\downarrow\downarrow}|^{2})e^{-\Gamma^{\downarrow}_{2}\tau_{2}}; \nonumber 
\\
P_{\rm{no,yes}}&=&-P_{\rm{no,no}} + (|c_{\uparrow\uparrow}|^{2} + 
|c_{\uparrow\downarrow}|^{2})e^{-\Gamma^{\uparrow}_{1}\tau_{1}} + \nonumber \\ & &
(|c_{\downarrow\uparrow}|^{2} + |c_{\downarrow\downarrow}|^{2})e^{-\Gamma^{\downarrow}_{1}\tau_{1}}; \nonumber 
\\
P_{\rm{yes,yes}}&=& 1 - P_{\rm{no,no}} - P_{\rm{yes,no}} - P_{\rm{yes,yes}}. \nonumber
\end{eqnarray}
These equations and the constraint $|c_{\uparrow\uparrow}|^{2} + |c_{\uparrow\downarrow}|^{2} + 
|c_{\downarrow\uparrow}|^{2} +
|c_{\downarrow\downarrow}|^{2} = 1$ are sufficient to determine the amplitudes of the two-qubit state.

In conclusion, we have investigated a problem of coupled two-level systems under the drive of a harmonic field. 
We have shown that the dynamics of this system is such that an effective tunable coupling, controlled by the 
intensity of the field, can be achieved, and as a result
coherent controllable transfer of population resulting in entanglement is possible.


The contribution of H.~T.~Pitk\"anen, who wrote the code for CNOT gates is gratefully acknowledged.
This work was supported by the Academy of Finland (Acad. Res. Fellowship no. 00857 and projects no. 7111994 and 
no. 7205476) and EU (SQUBIT-2 IST-1999-10673, and Marie Curie Fellowship HPMF-CT-2002-01893).
The author wishes to thank Y.~Makhlin, M.~Bremner, and W.~D\"ur for useful discussions, J.~J.~Virtanen and 
V.~Bergholm for providing us with a copy of their program, and  Prof.~P.~Zoller and the Austrian Academy of 
Sciences for supporting a
research visit at IQOQI Innsbruck, during which this paper was finalized.

\end{document}